\documentclass[%
aip,
superscriptaddress,
amsmath,amssymb,
reprint,%
]{revtex4-2}

\usepackage{graphicx}
\usepackage{dcolumn}
\usepackage{bm}

\usepackage[utf8]{inputenc}
\usepackage[T1]{fontenc}
\usepackage{booktabs, array, mathptmx, float, tabularx, booktabs, lipsum, amsmath,multirow}
\usepackage{siunitx, xcolor}
\usepackage[version=4]{mhchem}

\usepackage{times}
\usepackage{amsmath}
\usepackage{mathptmx}
\DeclareMathAlphabet{\mathcal}{OMS}{cmsy}{m}{n}
\DeclareSymbolFont{largesymbols}{OMX}{cmex}{m}{n}
\usepackage{mathrsfs}
\usepackage{epstopdf}
\usepackage[colorlinks=true, letterpaper=ture, pdfstartview=FitV, linkcolor=blue, citecolor=blue, urlcolor=blue]{hyperref}
\usepackage{dcolumn}

\begin{document}
	
	\title{Global Rotation of Skyrmion Bags under Vertical Microwave Fields}
	\author{Lan Bo}
	\affiliation{Department of Applied Physics, Waseda University, Okubo, Shinjuku-ku, Tokyo 169-8555, Japan}
	\affiliation{Key Laboratory for Anisotropy and Texture of Materials (MOE), School of Materials Science and Engineering, Northeastern University, Shenyang 110819, China}
	\author{Rongzhi Zhao}
	\email[Corresponding E-mail: ]{zhaorz@hdu.edu.cn}
	\affiliation{Institute of Advanced Magnetic Materials, College of Materials and Environmental Engineering, Hangzhou Dianzi University, Hangzhou 310012, China}
	\author{Xichao Zhang}
	\affiliation{Department of Applied Physics, Waseda University, Okubo, Shinjuku-ku, Tokyo 169-8555, Japan}
	\author{Masahito Mochizuki}
	\email[Corresponding E-mail: ]{masa\_mochizuki@waseda.jp}
	\affiliation{Department of Applied Physics, Waseda University, Okubo, Shinjuku-ku, Tokyo 169-8555, Japan}
	\author{Xuefeng Zhang}
	\affiliation{Key Laboratory for Anisotropy and Texture of Materials (MOE), School of Materials Science and Engineering, Northeastern University, Shenyang 110819, China}
	\affiliation{Institute of Advanced Magnetic Materials, College of Materials and Environmental Engineering, Hangzhou Dianzi University, Hangzhou 310012, China}
	\date{\today}
	
	\begin{abstract}
		Magnetic skyrmion bags are composite topological spin textures with arbitrary topological charges. Here, we computationally study the transient rotational motion of skyrmion bags, which is characterized by a global rotation of the inner skyrmions around the central point. Distinct from conventional rotational modes found in skyrmions, the observed rotation is a forced motion associated with the breathing mode induced solely by vertical microwave fields. The driving force behind this rotation originates from the interactions between outer and inner skyrmions, with the angular velocity determined by the phase difference resulting from their asynchronous breathing behaviors. It is also found that skyrmion bags with larger skyrmion numbers are more conducive to the occurrence of the rotation. Our results are useful for understanding the cluster dynamics of complex topological spin textures driven by dynamic fields.
		
	\end{abstract}
	
	\maketitle
	
	\section{Introduction}
	Magnetic skyrmions are topologically nontrivial spin textures and have garnered substantial attention and interest in various of fields ranging from fundamental physics to materials science.\cite{nagaosa2013topological,mochizuki2015dynamical,fert2017magnetic,leliaert2018fast,zhang2020skyrmion,gobel2021beyond,bo2022micromagnetic,reichhardt2022statics} Notably, recent advances in the exploration of localized spin configurations lead to the expansion of the skyrmion family. \cite{roessler2006spontaneous,finocchio2016magnetic,jiang2017skyrmions,kanazawa2017noncentrosymmetric,everschor2018perspective,back20202020,fujishiro2020engineering} For example, elementary particle-like skyrmions may be assembled together to form composite structures known as skyrmion bags,\cite{foster2019two} which consist of an outer skyrmion boundary (i.e., a circular domain wall) enveloping multiple smaller nested skyrmions. Skyrmion bags were initially proposed in 2019,\cite{foster2019two,rybakov2019chiral} and have since been experimentally detected in diverse materials, such as B20-type chiral magnets \cite{tang2021magnetic} and van der Waals magnets.\cite{powalla2023seeding} They have also been verified as transition states in the transformation processes involving skyrmions and skyrmioniums,\cite{yang2023reversible} as well as in the generation of skyrmion-antiskyrmion pairs.\cite{zheng2022skyrmion} The arbitrary topological charges endow skyrmion bags with the potential to enable high-density and multi-data information encoding,\cite{foster2019two,rybakov2019chiral} thereby attracting extensive theoretical investigations, including their existence and stability,\cite{kind2020existence} controllable creation,\cite{bo2023controllable} and dynamics driven by spin-orbit torques,\cite{zeng2020dynamics} spin-transfer torques,\cite{kind2021magnetic} anisotropy gradient,\cite{zeng2022skyrmion} and voltage gates.\cite{chen2022voltage,zhang2022high} Moreover, skyrmion bags could also serve as compelling evidence to physically explicate the particle-continuum duality \cite{wang2023particle} and extraordinary diversity \cite{kuchkin2020magnetic} of the skyrmion solutions. Therefore, the study of skyrmion bags is of academic significance and may lead to practical applications.
	
	The spin excitation of topological spin textures under microwave fields is an important issue in the control and manipulation of their dynamics. Dynamic responses in both skyrmion crystals \cite{mochizuki2012spin,onose2012observation} and individual skyrmion \cite{kim2014breathing,liu2018shape} have been investigated, where three eigenmodes have been identified. Specifically, these modes include a breathing mode under out-of-plane microwave fields, as well as clockwise and anticlockwise rotational modes under in-plane microwave fields.\cite{mochizuki2012spin,onose2012observation,kim2014breathing,liu2018shape} These resonance effects suggest an opportunity for the design and development of skyrmion-based microwave applications including microwave detectors \cite{mochizuki2013magnetoelectric,finocchio2015skyrmion} and nano-oscillators.\cite{liu2015dynamical,navau2016analytical} Furthermore, subsequent works have extended to other skyrmionic textures, such as antiskyrmion,\cite{raeliarijaona2018boundary,mckeever2019characterizing} skyrmionium,\cite{vigo2021spin} bimeron,\cite{wang2022field} and hopfion,\cite{liu2018binding,bo2021spin} whose individual spin eigenmodes can serve as unique fingerprints to enable their differentiation.\cite{lonsky2020dynamic}
	
	The spin excitation of skyrmion bags under out-of-plane \cite{zeng2022spin} and in-plane microwave fields \cite{li2023plane} have been explored very recently, with various oscillation modes identified. These works involved skyrmion bags with small skyrmion numbers and high central symmetry, constrained within a nanodisk. In this paper, we find that a global rotation of the skyrmion bags can be driven by a microwave field, where skyrmion bags with notably large skyrmion number is considered in an open boundary framework. The initial skyrmion bags inherently adopt a hexagonal spatial arrangement similar to the skyrmion crystals.\cite{matsuki2023thermoelectric} Different from the conventional skyrmion rotational modes induced by in-plane microwave fields \cite{mochizuki2012spin,onose2012observation,kim2014breathing,liu2018shape}, the observed transient rotation is believed to be a forced motion affiliated to the high-frequency response mode purely triggered by vertical microwave fields. 
	
	\begin{figure*}[t]
		\includegraphics[width=1\linewidth]{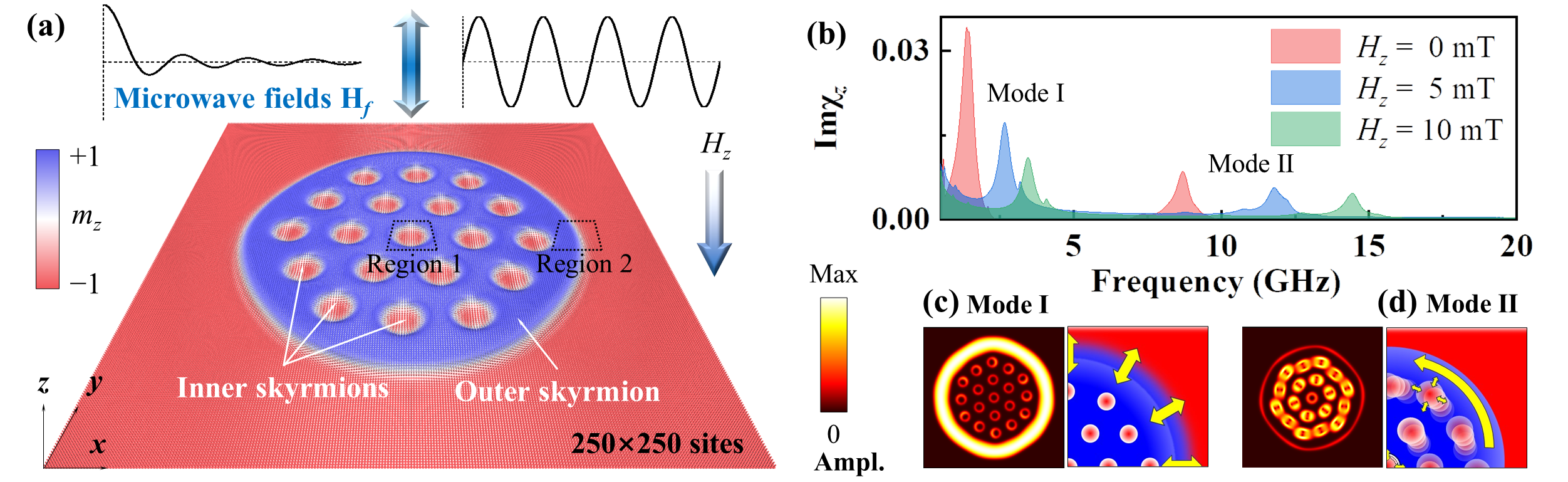}
		\caption{(a) Schematic of the micromagnetic model: a 2D magnetic film comprising a grid of $250\times250$ sites. The initial skyrmion bag consists of an outer skyrmion and nineteen inner skyrmions with opposite polarity. Regions 1 and 2 are squared areas each spanning dimensions of $30\times30$ sites, located at the central or along the border areas of the skyrmion bags, respectively. A uniform external field $-H_z$ is applied across the entire space. Microwave magnetic fields with sinc-function and sin-function profiles are shown as insets. The color bar of magnetization component $m_z$ is also applicable to all subsequent spin configurations. (b) Imaginary part of the susceptibility spectrum obtained after applying	a sinc-function field. Two resonance peaks under  $H_z=0$ are located at $f_{\rm\uppercase\expandafter{\romannumeral1}}=1.4 \ \rm GHz$ and $f_{\rm\uppercase\expandafter{\romannumeral2}}=8.7 $. Insets: schematic illustrations of the dynamic changes for the two spin excitation modes. (c) (d) The spatial FFT amplitude distributions of the z-component magnetization $m_z$. The color bar indicates the intensity of the oscillation amplitude.
		}\label{1} 
	\end{figure*}
	
	\section{Model and methodology}
	As illustrated schematically in Fig.~\ref{1} (a), the model considered in this work is a two-dimensional (2D) square lattice consisting of $N=250\times250$ sites, designed to simulate a magnetic thin film. A skyrmion bag with skyrmion number $\left| Q \right|=18$ is set as the initial spin configuration after full relaxation.\cite{bo2023controllable} Here, the skyrmion number is defined as \cite{nagaosa2013topological}
	\begin{align}
		{Q}=\frac{1}{4\pi}\int{\bf m}\cdot(\partial_{x}{\bf m}\times\partial_{y}{\bf m}) \,{\rm d}x\,{\rm d}y
		,\label{1}
	\end{align}
	where ${\bf m}$ is the normalized magnetization vector, and $n=\left| Q \right|+1$ is the actual number of the inner skyrmions because that the outer skyrmion carries $Q=1$. The diameter of the outer skyrmion is around 180 sites, which avoids the potential influence of finite geometric constraints. The average energy density of this system is given by 
	\begin{align}
		&\epsilon=A(\nabla {\bf m})^2+D\left[m_{z}(\nabla \cdot {\bf m})-({\bf m} \cdot \nabla) m_z\right]
		\nonumber\\
		&\ \  -K({\bf n}\cdot{\bf m})^2-\mu_0 M_{\rm s} m_z H_z-\frac{1}{2} \mu_0 M_{\rm s} {\bf m} \cdot {\bf H}_{\rm dm}
		,\label{2}				
	\end{align}
	where $A$, $D$, and $K$ are the Heisenberg exchange, interfacial Dzyaloshinskii-Moriya interaction (DMI), and perpendicular anisotropy constants, respectively. The vector ${\bf n}$ is the unit surface normal vector, $\mu_0$ is the vacuum permeability, $M_{\rm s}$ is the saturation magnetization, $H_z$ is the strength of the static applied magnetic ﬁeld along the $-z$ direction, and ${\bf H}_{\rm{dm}}$ is the demagnetizing field. To describe the dynamic behaviors of the skyrmion bags, we employ the finite-difference micromagnetic solver MUMAX3,\cite{vansteenkiste2014design} for the integration of the Landau-Lifshitz-Gilbert equation
	\begin{align}
		{\partial_t {\bf m}}=-\gamma_{0} {\bf m} \times {\bf h}_{\rm eff}+\alpha({\bf m} \times \partial_t {\bf m})
		,\label{3}		
	\end{align}
	where $\gamma_{0}$ is the absolute gyromagnetic ratio, $\alpha$ is the Gilbert damping constant, and $ {\bf h}_{\rm eff}=-(\delta \epsilon / \delta {\bf m}) /(\mu_0 M_{\rm s})$ is the effective field. The key input parameters are derived from Co/Pt films in real experiments \cite{sampaio2013nucleation,metaxas2007creep} and some previous theoretical works:\cite{kind2020existence,bo2023controllable,zeng2020dynamics}
	$M_{\rm s}=5.8\times10^{5}\, \rm{A}/\rm{m}$,
	$A=1.5\times10^{-11}\ \rm{J}/\rm{m}$,
	$D=3.5\times10^{-3}\ \rm{J}/\rm{m}^2$,
	$K=8.0\times10^5\ \rm{J}/\rm{m}^3$, and $\alpha=0.01-0.05$ depending on various situations.
	
	\section{Results and discussion}
	Our investigation starts by identifying the spin excitation modes of the skyrmion bags under vertical microwave fields. We apply an alternating current (AC) magnetic field ${\bf H}_f^{*}=[\,0,\, 0,\, H_{\rm 0}^{*}{\rm sin}(2\pi f t)/(2\pi f t)] $, with the amplitude $H_{\rm 0}^{*}=1\ \rm mT$ and the cutoff frequency $f=50\ \rm GHz$, and its profile can be found in the inset of Fig.~\ref{1} (a). The spin excitation spectra under different $H_z$ are given in Fig.~\ref{1} (b), where the imaginary part of the dynamical susceptibility, $\rm Im\chi_{\it z}$, is calculated from the fast Fourier transformation (FFT) of magnetization component $m_z(t)=(1/N)\ \Sigma_{i=1}^N m_{z,i}(t)$. Note that we present only the range within 20 GHz since there are no apparent resonance peaks observed beyond this range. Two notable resonance peaks have been identified, and these peaks exhibit a shift towards higher frequency intervals with the increase of $H_z$, while simultaneously weakening in strength. To further trace the spin dynamics of each mode, we consider the case where $H_z=0$ and apply sin-function AC fields ${\bf H}_{f}=\left[\,0,\, 0,\, H_{\rm 0}{\rm sin}(2\pi f_i\,t)\right] $, with the amplitude $H_{\rm 0}=15\ \rm mT$ and two eigenfrequencies $f_{\rm\uppercase\expandafter{\romannumeral1}}=1.4 \ \rm GHz$, $f_{\rm\uppercase\expandafter{\romannumeral2}}=8.7 \ \rm GHz$. The profile of such microwave fields is also shown in the inset of Fig.~\ref{1} (a). Figures~\ref{1} (c) (d) depict the amplitude distributions of the oscillation obtained through FFT for each spatial point. It is revealed that mode \rm\uppercase\expandafter{\romannumeral1} primarily corresponds to the breathing mode of the outer skyrmion, whereas mode \rm\uppercase\expandafter{\romannumeral2} is mainly the breathing mode of the inner skyrmions, accompanied by a collective rotational motion. The corresponding schematic illustrations of the spin dynamics for each mode are also shown at the side in Fig.~\ref{1} (c) (d). Interestingly, in previous studies concerning skyrmion excitations,\cite{mochizuki2012spin,onose2012observation,kim2014breathing} there have been no instances where a vertical magnetic field induced a rotational mode. So our following investigation will focus on characterizing and analyzing this rotational motion.
	
	\begin{figure}[t]
		\includegraphics[width=1\linewidth]{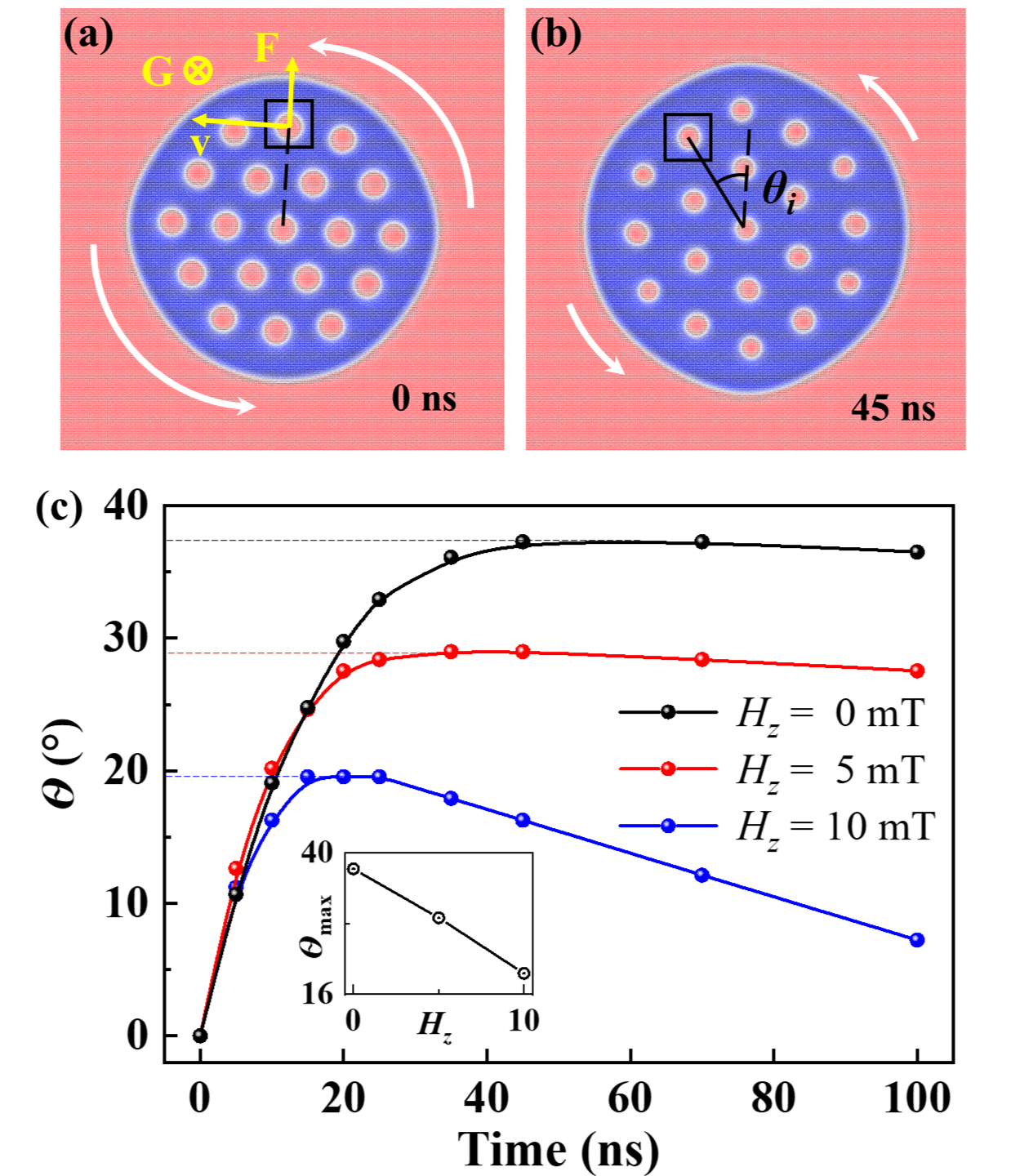}
		\caption{Selected snapshots of dynamic spin textures for a microwave-irradiated skyrmion bag with rotating inner skyrmions at (a) $t=0$ ns, and (b) $t=45$ ns. The rotational direction is indicated by white arrows. The schematic of Magnus force is indicated by yellow arrows. $\theta_{i}$ is the rotation angle of an inner skyrmion. (c) Time profiles of the rotation angle $\varTheta$ for various $H_z$. Inset: Variation of the maximum value of the rotation angle $\varTheta_{max}$ versus $H_z$.}\label{2} 
	\end{figure}
	
	\begin{figure}[!htb]
		\includegraphics[width=1\linewidth]{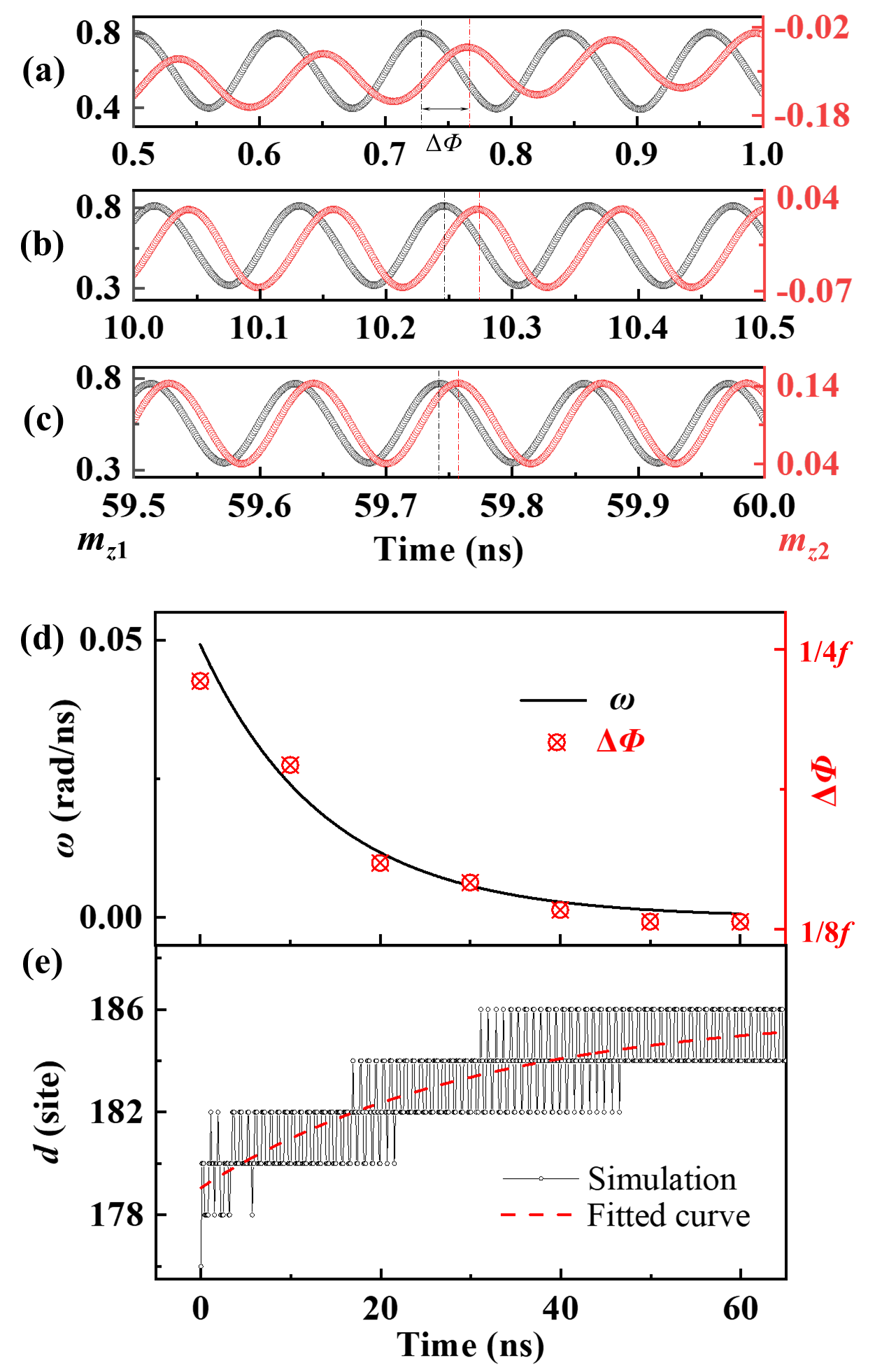}
		\caption{Variations of the $z$-component magnetizations in Region 1 and 2, $m_{z1}$ and $m_{z2}$, plotted as a function of time within 0.5 ns, for three distinct intervals: (a) Initial stage spanning from 0.5 ns to 1.0 ns; (b) Intermediate stage spanning from 10.0 ns to 10.5 ns; (c) Stable stage spanning from 59.5 ns to 60.0 ns. $\Delta \varPhi$ represents the phase difference between $m_{z1}(t)$ and $m_{z2}(t)$. (d) Time-dependent variations of the angular velocity  $\omega$ (black curve) and the phase difference $\Delta \varPhi$ (red data points). (e) Variation of the outer skyrmion diameter $d$ versus time. The black line with markers represents the simulated results, while the red dashed line corresponds to the fitted curve.}\label{3} 
	\end{figure}

	Now we undertake the characterization of the rotational dynamics of the skyrmion bags. Figure~\ref{2} (a) (b) present the snapshots of the dynamic spin configurations throughout the rotation process under $H_z=0$. It is evident that the inner skyrmions collectively engage in an anticlockwise rotation around the central point. We trace a marked inner skyrmion as an example, which, at a given moment $t$, will rotate by an angle denoted as $\theta_{i}(t)$, from its initial position. Here, $i$ is the index of inner skyrmions. As shown in the inset of Fig.~\ref{2} (a), because the shape of inner skyrmions remain unchanged, the direction of rotation can be determined by the Thiele framwork \cite{jiang2017direct}  ${\bf G}\times{\bf v}\sim{\bf F}$, where ${{\bf G}=(0, 0, 4\pi Q)}$ is the gyromagnetic coupling vector, ${\bf v}$ is the skyrmion velocity, and ${\bf F}$ is the interaction force from the outer skyrmion boundary. This equation indicates that the moving direction of skyrmion $(\parallel{\bf v} \propto {\bf F}\times{\bf G})$ is perpendicular to ${\bf F}$. More specifically, because the inner skyrmions have $Q=-1$, the force from the outer skyrmion boundary drives their motion in the azimuth direction of anticlockwise (clockwise) sense, when ${\bf F}$ is attractive (repulsive). Then, we introduce the rotation angle of the skyrmion bags $\varTheta(t)=(1/n)\ \Sigma_{i=1}^n \theta_{i}(t)$, with $\theta_{i}(t)$ being the rotation angle of each inner skyrmion. The variation of $\varTheta$ versus time under different $H_z$ is shown in Fig.~\ref{2} (c). It can be seen that the rotational motion is a transient phenomenon. Specifically, as time progresses, $\varTheta$ exhibits a rapid increase, followed by a gradual decline. For $H_z=0$, the maximum value of the rotation angle $\varTheta_{max}$ reaches up to about 40 degrees. As $H_z$ increases, $\varTheta$ tends to reach the maximum value at an earlier stage, and $\varTheta_{max}$ undergoes a gradual reduction, as shown in the inset of Fig.~\ref{2} (c). We also find that the applied static field $H_z$ enhances the decay in $\varTheta$. This is due to the fact that as $H_z$ increases, the outer skyrmion tends to shrink, and the inner skyrmions are subject to a repulsive force toward the center, leading to rotation in the opposite direction. The magnitude of the decay in $\varTheta$ is minimal when $H_z=0$, even as time extends to 100 ns. So the case of zero static field is exclusively considered in the following studies.

	We next aim at analyzing the physical origin of the rotational motion. As shown in Fig.~\ref{1}, two distinct regions are defined in the 2D space, each with $30\times30$ sites. Region 1 completely covers the central inner skyrmion, and Region 2 covers part of the boundary of the outer skyrmion. The time-dependent variations of $z$-component magnetizations within the two regions (defined as $m_{z1}$ and $m_{z2}$) are extracted, which respectively reflect the breathing behaviors of inner and outer skyrmions in response to the vertical microwave fields. We choose three representative time intervals: the initial stage from 0.5 to 1.0 ns, the intermediate stage from 10.0 to 10.5 ns, and the stable stage from 59.5 to 60.0 ns. Here, we concentrate on the time frame within 60 ns. This decision is based on the observation from Fig.~\ref{2}, which indicates that the rotation angle reaches its maximum at around 45 ns, suggesting that a 60 ns window is sufficient for our analysis. In Fig.~\ref{3} (a)-(c), the variations of $m_{z1}$ and $m_{z2}$ are plotted as a function of time for these selected stages. Figure~\ref{3} (a) shows distinct asynchronous oscillations, where the amplitude of the oscillation curve for $m_{z1}$ consistently remains stable, while the amplitude of $m_{z2}$ appears weaker and unstable. We define the phase difference $\Delta \varPhi$ as the time difference between the moments at which the fluctuations of $m_{z1}$ and $m_{z2}$ reach their peaks within one function period. It is obvious that $\Delta \varPhi$ becomes smaller as time increases. At this point, we put forward a conjecture that the transient rotational motion of skyrmion bags results from the asynchrony in the breathing behaviors between the inner skyrmions and the outer skyrmion.
	
	To verify the above hypothesis, we calculate the angular velocity $\omega$ of the rotation, which is the derivative of the rotation angle with respect to time, defined as $\omega=d\varTheta/dt$. We then compare the time profile of $\omega$ with that of the phase difference $\Delta \varPhi(t)$, as shown in Fig.~\ref{3} (d). From the consistency of the curve and the scatter points, we can find a significant positive correlation between $\omega$ and $\Delta \varPhi$, which both decrease from an initial value and gradually stabilize. This corroborates the conclusion that the rotation is induced by the asynchrony in breathing behaviors, and further demonstrates that the velocity of the rotational motion is determined by $\Delta \varPhi$. In order to investigate the underlying mechanism of the phase difference, we extracted the diameter of the outer skyrmion $d$ as a function of time during the rotation, as shown in Fig.~\ref{3} (e). The simulation results reveal that $d$ gradually increases during the oscillation and converge to a constant. This increasing trend is fitted by an exponential function with horizontal asymptote $d=a-b*c^t$, where $a=186.0$, $b=6.88$ and $c=0.97$. Regarding this phenomenon, we offer the following explanation: in this response mode, the breathing of inner skyrmions is spontaneous, while the breathing of the outer skyrmion is induced by the former. Hence, as time progresses, the phases of their oscillations trend to gradually converge. Furthermore, when inner skyrmions start continuous breathing from the initial static state, they exert an interaction potential that encourages expansion on the passively oscillating outer skyrmion, resulting in a slight increase in $d$. Consequently, the reaction of this potential becomes the driving force for the forced rotation of skyrmion bags.
	
	\begin{figure}[t]
		\includegraphics[width=1\linewidth]{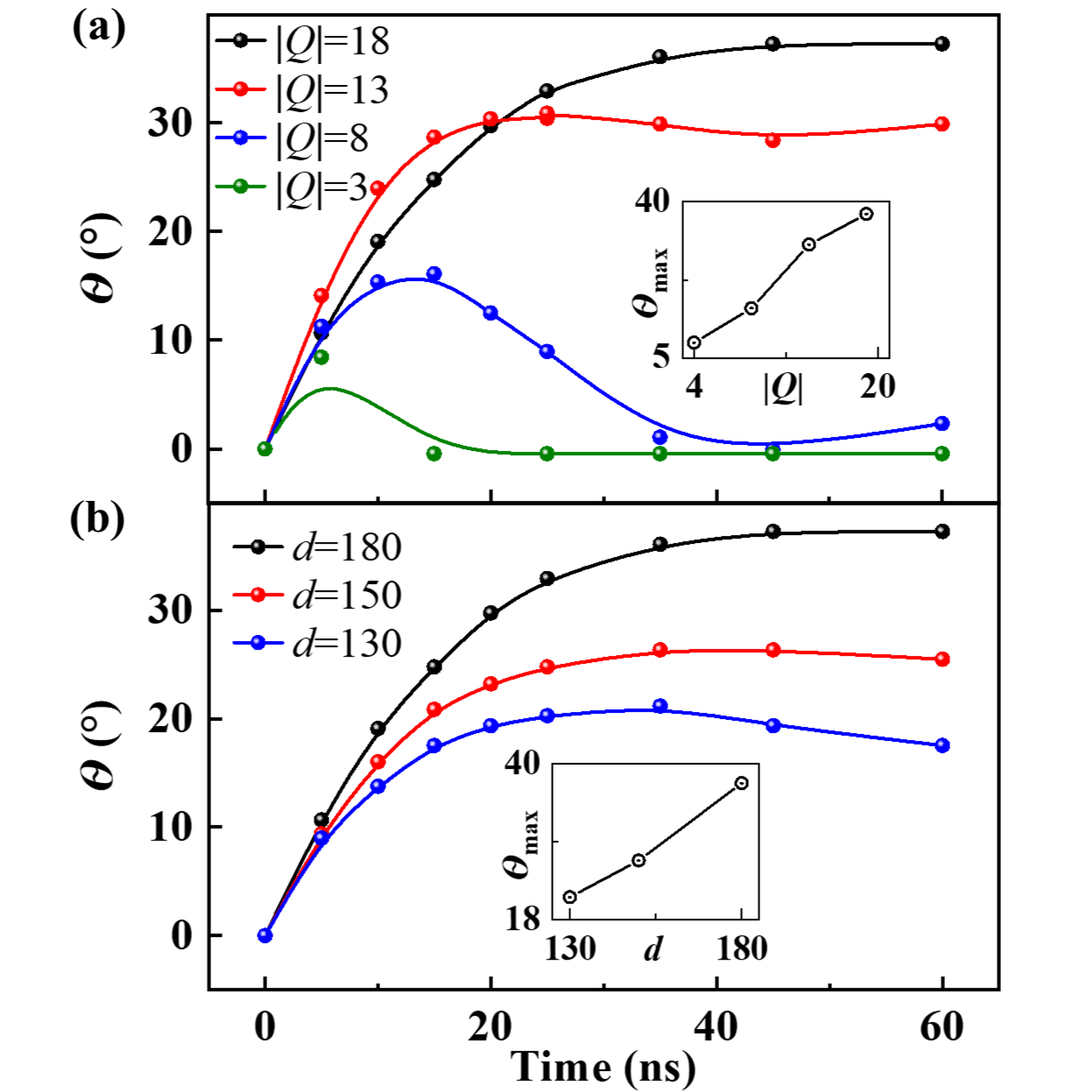}
		\caption{(a) The rotation angle $\varTheta$ plotted as a function of time for various skyrmion numbers $Q$. Inset: the maximum value of rotation angle $\varTheta_{max}$ for the four diffetenr $Q$. (b) The rotation angle $\varTheta$ plotted as a function of time for various outer skyrmion diameter $d$. Inset: the maximum value of rotation angle $\varTheta_{max}$ for the three diffetenr $d$.}\label{4} 
	\end{figure}
	
	Finally, we discuss the rotation behavior of skyrmion bags with different skyrmion numbers, where $\left| Q \right|=13$, 8, and 3 are selected for comparison. After applying ${\bf H}_f^{*}$ to find their resonance frequencies, we excite their high-frequency mode separately and record the time-dependent variations of $\varTheta$, as shown in Fig.~\ref{4} (a). Clearly, the variations can be categorized into two distinct trends. For skyrmion bags with $\left| Q \right|=18$ and 13, $\varTheta$ exhibits a slight decrease or fluctuation after reaching its maximum, depending on the specific symmetry of the spin configurations; whereas for skyrmion bags with $\left| Q \right|=8$ and 3, $\varTheta$ rapidly returns to zero after reaching its maximum, indicating a reverse rotation back to the initial position. We further extract the maximum value of rotation angle $\varTheta_{max}$, for the four situations, as shown in the inset of Fig.~\ref{4} (a). It can be summarized that for a smaller $\left| Q \right|$ of skyrmion bags, the achievable $\varTheta_{max}$ tends to be smaller. This is because when there are more inner skyrmions, the overall oscillation generated by their breathing mode is stronger, resulting in a larger driving force to maintain the rotation. The results indicates that the rotation is more likely to occur for skyrmion bags with larger $\left| Q \right|$, while it is less likely to be observed in skyrmion bags with smaller $\left| Q \right|$. Interestingly, we noticed that as $\left| Q \right|$ decreases, the diameter of the outer skyrmion $d$ also decreases at the same time. In order to verify whether the above conclusion is attributed by changes in $\left| Q \right|$ or $d$, we conducted additional simulations by proportionally reducing $d$ through adjusting the DMI constant. The impact of $d$ on the rotational behavior is illustrated in Fig.~\ref{4} (b). The reduction of $d$ does weaken the rotation, indicating that, for a constant $\left| Q \right|$, the rotational behavior is more pronounced in larger-sized skyrmion bags. However, the decrease in $d$ does not lead to a rapid decay similar to that observed for smaller $\left| Q \right|$. Because $d$ is approximately 130 sites when $\left| Q \right|=8$, we can compare the results of $\left| Q \right|=8$ and $d=130$ to confirm this point.
	
	\section{Conclusion and Prospect}
	In conclusion, we have investigated a global rotational motion of skyrmion bags in response to microwave magnetic fields. This behavior is specific to the high-frequency mode, and is characterized by the inner skyrmions rotating cohesively around the central point. It has been proven that the angular velocity of the rotation is directly related to the phase difference between the oscillations of inner and outer skyrmions. And the driving force of the rotation arises from the interaction potential of the outer skyrmion boundary acting on the inner skyrmions. It is also found that the rotation is more prone to occur in skyrmion bags with larger skyrmion numbers. 
	
	Although the rotation is a transient dynamics, it enables the possibility of manipulation of skyrmion bags by activating the eigenmode. Specially, it induces both spatial and temporal variations in magnetization, providing potential to generate rich spintronic phenomena and applications such as spinmotive force \cite{matsuki2023thermoelectric} and programmable logic device.\cite{yan2021skyrmion} In a real system, the rotation may be influenced by temperture, but we believe that under small thermal fluctuations, the skyrmion bag will experience some distortion without undergoing fundamental changes in its rotational behavior.\cite{matsuki2023thermoelectric} Our findings hold significance for the microwave-controlled dynamics of skyrmion bags, and may extend to other topological configurations featuring multiple domains.

	\begin{acknowledgments}
		X.C.Z. and M.M. acknowledge support by the JST CREST (Project No. JPMJCR20T1), the JSPS KAKENHI (Grants No. JP20H00337 and No. JP23H04522), and the Waseda University Grant for Special Research Projects (Grant No. 2023C-140). R.Z.Z and X.F.Z. acknowledge support by the National Science Fund for Distinguished Young Scholars (Grant No. 52225312), the National Natural Science Foundation of China (Grant No. U1908220), and the Key Research and Development Program of Zhejiang Province (Grant No. 2021C01033). L.B. thanks the financial support by China Scholarship Council (Award No. 202206080023).
	\end{acknowledgments}
	
	\section*{AUTHOR DECLARATIONS}
	The authors have no conflicts to disclose.
	
	\section*{Data Availability Statement}
	The data that support the findings of this study are available from the corresponding authors upon reasonable request.

	\bibliography{SB_rotation_revised}
\end{document}